\documentclass[aps,prl,reprint,superscriptaddress,longbibliography,showkeys,english]{revtex4-1}

\usepackage[T1]{fontenc}
\usepackage{lmodern}
\usepackage{color}
\usepackage{babel}
\usepackage{amsmath}
\usepackage{amssymb}
\usepackage{graphicx}

\RequirePackage[colorlinks=true,linkcolor=blue]{hyperref}
\usepackage{natbib}
 
\usepackage{babel}

\usepackage{physics}
\usepackage{xcolor}
\makeatother

\begin{document}

\title{Flopping-mode electric dipole spin resonance}
\author{X.~Croot}
\affiliation{Department of Physics, Princeton University, Princeton, New Jersey 08544, USA}
\author{X.~Mi}
\affiliation{Department of Physics, Princeton University, Princeton, New Jersey 08544, USA}
\author{S.~Putz}
\affiliation{Department of Physics, Princeton University, Princeton, New Jersey 08544, USA}
\author{M.~Benito}
\affiliation{Department of Physics, University of Konstanz, D-78457 Konstanz, Germany}
\author{F.~Borjans}
\affiliation{Department of Physics, Princeton University, Princeton, New Jersey 08544, USA}
\author{G.~Burkard}
\affiliation{Department of Physics, University of Konstanz, D-78457 Konstanz, Germany}
\author{J.~R.~Petta}
\affiliation{Department of Physics, Princeton University, Princeton, New Jersey 08544, USA}

\date{\today}

\begin{abstract}
Traditional approaches to controlling single spins in quantum dots require the generation of large electromagnetic fields to drive many Rabi oscillations within the spin coherence time. We demonstrate ``flopping-mode'' electric dipole spin resonance, where an electron is electrically driven in a Si/SiGe double quantum dot in the presence of a large magnetic field gradient. At zero detuning, charge delocalization across the double quantum dot enhances coupling to the drive field and enables low power electric dipole spin resonance. Through dispersive measurements of the single electron spin state, we demonstrate a nearly three order of magnitude improvement in driving efficiency using flopping-mode resonance, which should facilitate low power spin control in quantum dot arrays. 
\end{abstract}

\maketitle

Recent advances in silicon spin qubits have bolstered their standing as a platform for scalable quantum information processing. As single-qubit gate fidelities exceed 99.9$\%$ \cite{Yoneda2018}, two-qubit gate fidelities improve \cite{Dzurak2015, Xue2018, Huang2018, Zajac2018, Watson2018}, and the field accelerates towards large multi-qubit arrays \cite{Meunier2018, Mills2018}, developing the tools necessary for efficient and scalable spin control is critical \cite{Hanson07}. While it is possible to implement single electron spin resonance in quantum dots (QDs) using ac magnetic fields \cite{Koppens2006}, the requisite high drive powers and associated heat loads are technically challenging and place limitations on attainable Rabi frequencies \cite{Takeda2016}. As spin systems are scaled beyond a few qubits, methods of spin control which minimize dissipation and reduce qubit crosstalk will be important for low temperature quantum information processing \cite{Hornibrook2015}. 

Electric dipole spin resonance (EDSR) is an alternative to conventional electron spin resonance. In EDSR, static gradient magnetic fields and oscillating electric fields are used to drive spin rotations \cite{Rashba03}. The origin of the effective magnetic field gradient varies across implementations: intrinsic spin-orbit coupling \cite{Golovach06,Flindt06,Nowack2007}, hyperfine coupling~\cite{Laird2007}, and $g$-factor modulation \cite{Kato2003} have been used to couple electric fields to spin states. The inhomogeneous magnetic field generated by a micromagnet \cite{Pioro-Ladriere2008,Kawakami2014} has been used to create a synthetic spin-orbit field for EDSR, enabling high fidelity control~\cite{Yoneda2018}. Conveniently, this magnetic field gradient gives rise to a spatially varying Zeeman splitting, enabling spins in neighboring QDs to be selectively addressed~\cite{Pioro-Ladriere2008,Obata2010,Nadj-Perge2010,Yoneda2014,Noiri2016,Takeda2016,Ito2018}. 

In this Letter, we demonstrate a novel mechanism for driving low-power, coherent spin rotations, which we call ``flopping-mode EDSR''. In conventional EDSR, the electric drive field couples to a charge trapped in a single quantum dot, leading to a relatively small electronic displacement \cite{Nowack2007}. We instead drive single spin rotations in a DQD close to zero detuning, $\epsilon$ = 0, where the electric field can force the electron to flop back and forth between the left and right dots, thereby sampling a larger variation in transverse magnetic field. We call this configuration the ``flopping-mode''.

Neglecting spin, the Hamiltonian describing a single electron trapped in a DQD is given by $H_0$ = $(\epsilon/2) \tau_z$ + $t_c \tau_x$, where $t_c$ is the interdot tunnel coupling and $\tau_i$ are the Pauli operators in position (L, R) space \cite{Hayashi03}. In the highly detuned regime of a DQD (with $|\epsilon|$$\gg$$t_c$), the electron is strongly localized in either the left $\ket{L}$ or right $\ket{R}$ dot, and the relevant orbital energy scale is $E_{\rm orb} \approx$ 3--5 meV. In contrast, when $\epsilon$ = 0 the charge delocalizes across the DQD leading to the formation of bonding and antibonding states $\ket{\mp}$ = $(\ket{R}\mp\ket{L})/\sqrt{2}$. Here the bonding-antibonding energy difference $2t_c$~$\approx$~20--40~$\mu$eV is dominant and the charge is much more susceptible to oscillating electric drive fields \cite{HuNori12, Kim2015}.

The application of a magnetic field results in Zeeman splitting of the spin states. When the Zeeman energy and tunnel splitting $2t_c$ at $\epsilon$~=~0 are comparable, the combination of a magnetic field gradient and the large electric dipole moment result in strong spin-charge hybridization. This allows electric fields to couple to spin indirectly via the charge ~\cite{Viennot15, Mi2018, Samkharadze2018}. We coherently manipulate a single-electron spin qubit in the flopping-mode regime and find that the power required to drive Rabi oscillations is almost three orders of magnitude less than in single dot EDSR. In addition to improving the spin sensitivity to electric fields, there is a ``sweet spot'' at $\epsilon$ = 0 where charge noise is suppressed \cite{Vion02}, leading to a four-fold improvement in qubit quality factor.

\begin{figure*}
\centering
\includegraphics[width=7in]{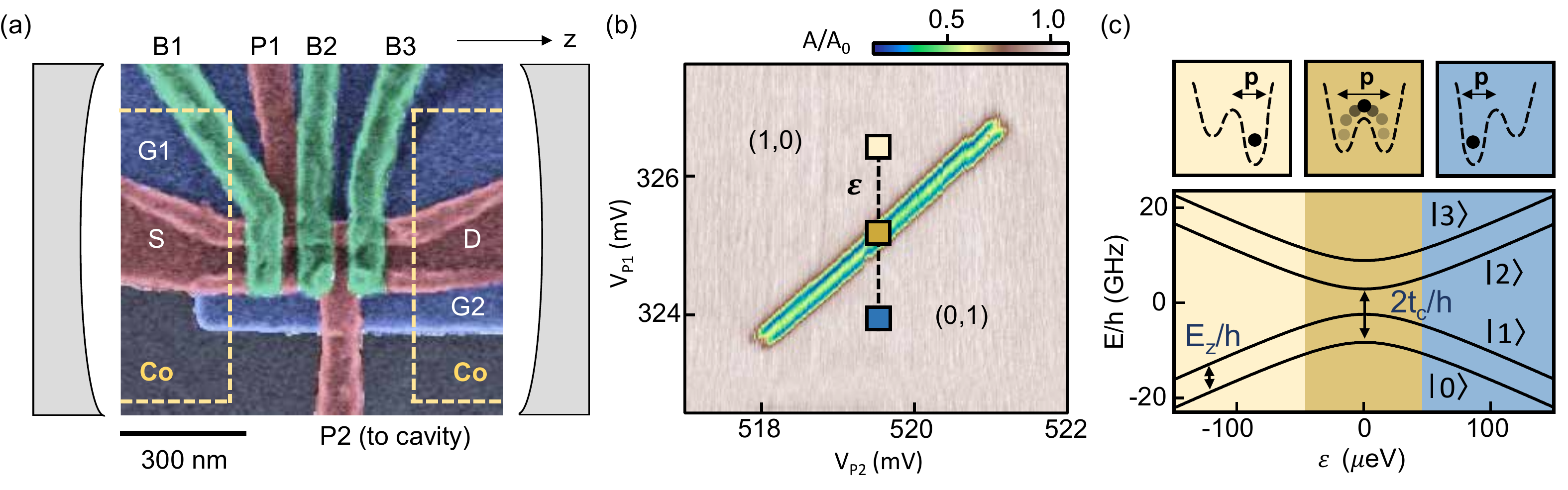}
\protect\caption{\label{fig:figure1} 
(a) False-color SEM image of the device. The locations of the cobalt micromagnets are indicated by the yellow dashed lines. Gate P2 of the DQD is galvanically connected to the center pin of the superconducting cavity. (b) Normalized cavity transmission amplitude $A/A_0$ as a function of the gate voltages $V_{\text{P}1}$ and $V_{\text{P}2}$ near the interdot charge transition for DQD2, with $2t_c/h = 4.9\,  \text{GHz}$. The dashed arrow denotes the DQD detuning parameter $\epsilon$. (c) Top: Schematic comparison of the far detuned single dot regime ($|\epsilon| \gg 2t_c$) and flopping-mode regime ($\epsilon \approx 0$). Charge hybridization near $\epsilon$ = 0 results in a large electric dipole moment $\vec{p}$ compared to the single dot regime. Bottom: Energy diagram calculated with $2t_c/h = 11.1\, \text{GHz}$, $B_z = 209.4\, \text{mT}$, $b_z = 0\, \text{mT}$ and $b_x = 15\, \text{mT}$.
}
\end{figure*}

The device consists of two single-electron natural-Si/SiGe DQDs (DQD1 and DQD2) that are embedded in a half-wavelength niobium superconducting cavity with resonance frequency $f_c$ = 5.846 GHz \cite{Mi2018}. A false-color scanning electron microscope (SEM) image of one of the DQDs is shown in Fig.~\ref{fig:figure1}(a). For the purposes of this experiment, only one DQD is active at a time. Electronic confinement is achieved using an overlapping aluminum gate stack \cite{Zajac2016}. For each DQD, the plunger gate P2 is connected to the center pin of the superconducting cavity, efficiently coupling the charge confined in the DQD to the electric field of the cavity.

Cobalt micromagnets [yellow dashed lines in Fig.~\ref{fig:figure1}(a)] generate a local magnetic field gradient with longitudinal and transverse components. The total magnetic field at the device $\vec{B}^{\rm tot}$ is the sum of the external magnetic field $\vec{B}^{\rm ext}$ applied in the z direction, and the stray field from the micromagnet $\vec{B}^{\rm M}$.  As a result, the total magnetic field $\vec{B}^{\rm tot}$ is a sensitive function of position in the DQD confinement potential. The micromagnet enables EDSR \cite{Pioro-Ladriere2008,Kawakami2014} and gives rise to spin-photon coupling, which is  utilized here solely for dispersive cavity readout of the spin state \cite{Petersson2012,Mi2018}.

We first probe the charge degree of freedom with $B^{\text{tot}}$ = 0 by measuring the cavity transmission amplitude $A/A_0$ in the single photon regime, as a function of the gate voltages $V_{\text{P}1}$ and $V_{\text{P}2}$  [see Fig.~\ref{fig:figure1}(b)]. These data are acquired near the (1,0)--(0,1) interdot charge transition with the probe frequency equal to the cavity frequency $f$ = $f_c$. Here ($n_L$, $n_R$) denotes the charge occupancy of the DQD, where $n_L(n_R)$ is the number of electrons in the left(right) dot. Around $\epsilon$ = 0, where the DQD is maximally polarizable, electric fields from the cavity result in charge dynamics within the DQD that load the superconducting cavity and reduce $A/A_0$ \cite{Petersson2012,Frey12}. Neglecting spin, the energy difference between the bonding and antibonding charge states is $\Omega(\epsilon)$ = $\sqrt{\epsilon^2 + 4t_c^2}$. In the case where $2t_c$ < $hf_c$, there will be two values of detuning where $\Omega(\epsilon)$ = $h f_c$ ($h$ is Planck's constant). Around these values of detuning the cavity response is substantial, as is evident from the data in Fig.~\ref{fig:figure1}(b) \cite{Frey12,Petersson2012}. Since the flopping-mode EDSR mechanism is based on charge motion in a magnetic field gradient, it will also be most effective near $\epsilon$ = 0, where the charge dipole moment $\vec{p}$ is largest [Fig.~\ref{fig:figure1}(c)].

In the presence of a micromagnet and external magnetic field, the Hamiltonian describing the one-electron DQD is \begin{equation}
H_s = H_0  + \frac{1}{2}g \mu_B \lbrack B_z \sigma_z + (b_x \sigma_x + b_z \sigma_z) \tau_z\rbrack,
\end{equation}
\noindent
where $\sigma_i$ are the Pauli operators in spin space, $B_z$ is the homogeneous magnetic field component in the z direction, $g$ is the electronic $g$-factor, and $\mu_B$ is the Bohr magneton \cite{Benito2017}.  In general, $\vec{B}^{\rm M}$ will generate longitudinal $b_z$ and transverse ($b_x$, $b_y$) gradients that will modify the energy level spectrum. We define 2$b_i$ ($i$ = $x$, $y$, $z$) as the difference in total magnetic field between the left and right dots in the $i$ direction.  Without loss of generality, we take $b_y$ = 0 in the remainder of the Letter \cite{Benito2019}.  The total magnetic field components at the left and right dots can be written as $B^{\rm tot}_{z}$ = $B_z$ $\pm$ $b_z$ and $B^{\rm tot}_{x}$ = $B_x$ $\pm$ $b_x$, where $B_{z(x)}$ is the homogeneous magnetic field in the z(x) direction. For our micromagnet design, we expect $B_x$ $\approx$ 0, but note that in the case where $B_x$ $\neq$ 0, the geometric coordinate system can always be rotated to satisfy $B_x$ = 0 \cite{Benito2019}. The Zeeman energy is given by $E_z$ = $g\mu_B B_z$.

Zeeman splitting of the bonding/antibonding states leads to the 4-level system shown in Fig.~\ref{fig:figure1}(c), where $\ket{0(1)} \equiv \ket{-,\downarrow(\uparrow)}$ refer to the spin-down and spin-up bonding states, and $\ket{2(3)} \equiv
\ket{+,\downarrow(\uparrow)}$ to the spin-down and spin-up antibonding states. Spin-preserving interdot tunnel coupling $t_c$ results in the anticrossings near $\epsilon$ = 0. The $b_z$ component gives rise to a spatially varying longitudinal field such that the energy splitting $E_{01}$ between the ground $\ket{0}$ and first excited state $\ket{1}$ can vary significantly in the far detuned limits, e.g.\ $\epsilon \gg 0$ versus $\epsilon \ll 0$. The transverse components of the micromagnet field hybridize the spin-orbital states near $\epsilon$ = 0 and lead to spin-photon coupling \cite{Benito2017,Mi2018}, which enables dispersive spin state readout \cite{Petersson2012}.

\begin{figure}[t]
\includegraphics[width=\columnwidth]{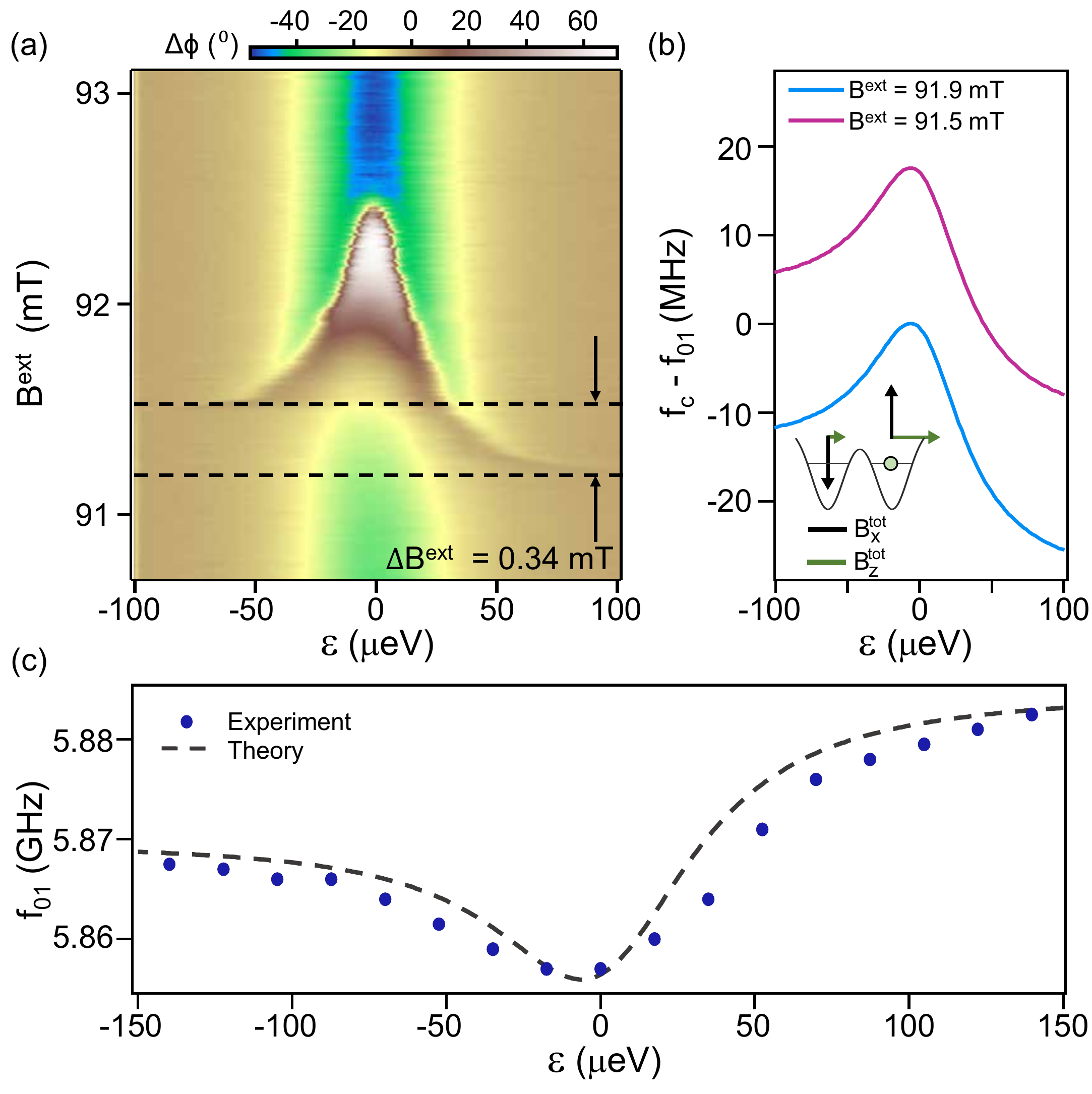}
\protect\caption{\label{fig:figure2} 
(a) Phase response of the cavity transmission $\Delta \phi$ as a function of $B^{\rm ext}$ and $\epsilon$ for DQD1 with $2t_c/h = 7.4\, \text{GHz}$.  Dashed lines show asymmetry in spin-cavity interactions at high and low detuning. The extracted $\Delta B^{\rm ext}$ is used to determine $b_z$. (b) Difference between the cavity frequency $f_c$ and spin transition frequency $f_{01}$ as a function of $\epsilon$ for $B^{\rm ext}$ = 91.9 mT (navy) and $B^{\rm ext}$ = 91.5 mT (purple). Inset: Cartoon of a DQD in the presence of spatially varying $B_x^{\rm tot}(B_z^{\rm tot})$ fields, not drawn to scale. (c) $f_{01}$ as a function of $\epsilon$ for DQD1, with $2t_c/h$ = 11.1 GHz, as extracted from time-domain Rabi oscillations. The dashed line shows a fit to theory with $2t_c/h$ = 11.1 GHz, $B_z = 209.4\, \text{mT}$, $b_x = 15\, \text{mT}$  and $b_z = 0.27\, \text{mT} $.
}
\end{figure}

In general, the electron spin resonance condition will be a function of detuning owing to the magnetic field gradients. To investigate the flopping-mode EDSR mechanism, we first map out the spin resonance condition by measuring the cavity phase shift $\Delta \phi$ as a function of $B^{\rm ext}$ and $\epsilon$ [Fig.~\ref{fig:figure2}(a)]. The funnel-shaped feature in Fig.~\ref{fig:figure2}(a) is a consequence of detuning-dependent charge hybridization and Zeeman physics \cite{Benito2017,Samkharadze2018} in the regime where $E_{01} \ll 2t_c$. At low $B^{\text{ext}}$, the spin transition is detuned from the cavity, but there is still a small phase response around $\epsilon$ = 0 due to the large electric dipole moment \cite{Frey12,Petersson2012}. At large detunings ($|\epsilon| \gg 2t_c$) the energy splitting $E_{01}$ is dominated by Zeeman physics. At small detunings, levels $\ket{1}$ and $\ket{2}$ hybridize due to transverse magnetic fields \cite{Benito2017}. This spin-charge hybridization pulls $E_{01}$ slightly below the Zeeman energy.  As a result, when $E_{01}$ is slightly less than $hf_c$ at $\epsilon$ = 0, there are two values of finite detuning for which $E_{01}$ is on resonance with the cavity, giving rise to the wings of the funnel-shaped feature that begins at $B^{\text{ext}}$ $\sim$ 91.2 mT. As $B^{\text{ext}}$ increases further the values of detuning that lead to resonance with the cavity shift closer to $\epsilon$ = 0. Eventually, at $B^{\text{\rm ext}}$ $\sim$ 91.9 mT, the two resonance conditions merge at $\epsilon$ = 0. Figure \ref{fig:figure2}(b) shows theoretical predictions for $f_c - f_{01}$ as a function of $\epsilon$ for $B^{\text{\rm ext}}$ = 91.5 mT and 91.9 mT, with $f_{01}$ $\equiv$ $E_{01}/h$.

With the electron spin resonance frequency $f_{01}$ now mapped out as a function of $B^{\rm ext}$ and $\epsilon$, we can drive coherent single spin rotations using flopping-mode EDSR. {At $\epsilon$ = 0, a microwave burst of frequency $f_s$ and duration $\tau_B$ is applied to gate P1 to drive coherent spin rotations. The final spin state is read out dispersively at $\epsilon$ = 0 by measuring the cavity phase response $\Delta \phi$ \cite{Mi2018}.

A typical flopping-mode dataset is shown in Fig.~\ref{fig:figure3}(a), where $\Delta \phi$ is plotted as a function of $f_s$ and $\tau_B$. As expected, the Rabi oscillation visibility is maximal when $f_s$ is resonant with $f_{01}$.  The spin transition frequency $f_{01}$ is plotted in Fig.~\ref{fig:figure2}(c) as a function of $\epsilon$. When $2t_c \gg E_z$, the lowest $f_{01}$ occurs near $\epsilon$  = 0 due to spin-charge hybridization, and $f_{01}$ increases as $|\epsilon|$ increases. The trends in these data are in general agreement with the data measured using microwave spectroscopy in Fig.~\ref{fig:figure2}(a). The asymmetry of the data in Figs.~\ref{fig:figure2}(a) and (c) about $\epsilon$ = 0  is due to the longitudinal gradient field $b_z$. Using the expression 2$b_z$ = (1+$\chi$)$\Delta$$B^{\text{ext}}$, where $\chi$ is the micromagnet magnetic susceptibility \cite{Mi2018} and $\Delta$$B^{\text{ext}}$ = 0.34 mT [from data in Fig.~\ref{fig:figure2}(a)], we find $b_z$ = 0.27 mT. We take this value and fit the data in Fig.~\ref{fig:figure2}(c), finding good agreement between experiment and our theoretical model.

\begin{figure}
\includegraphics[width=\columnwidth]{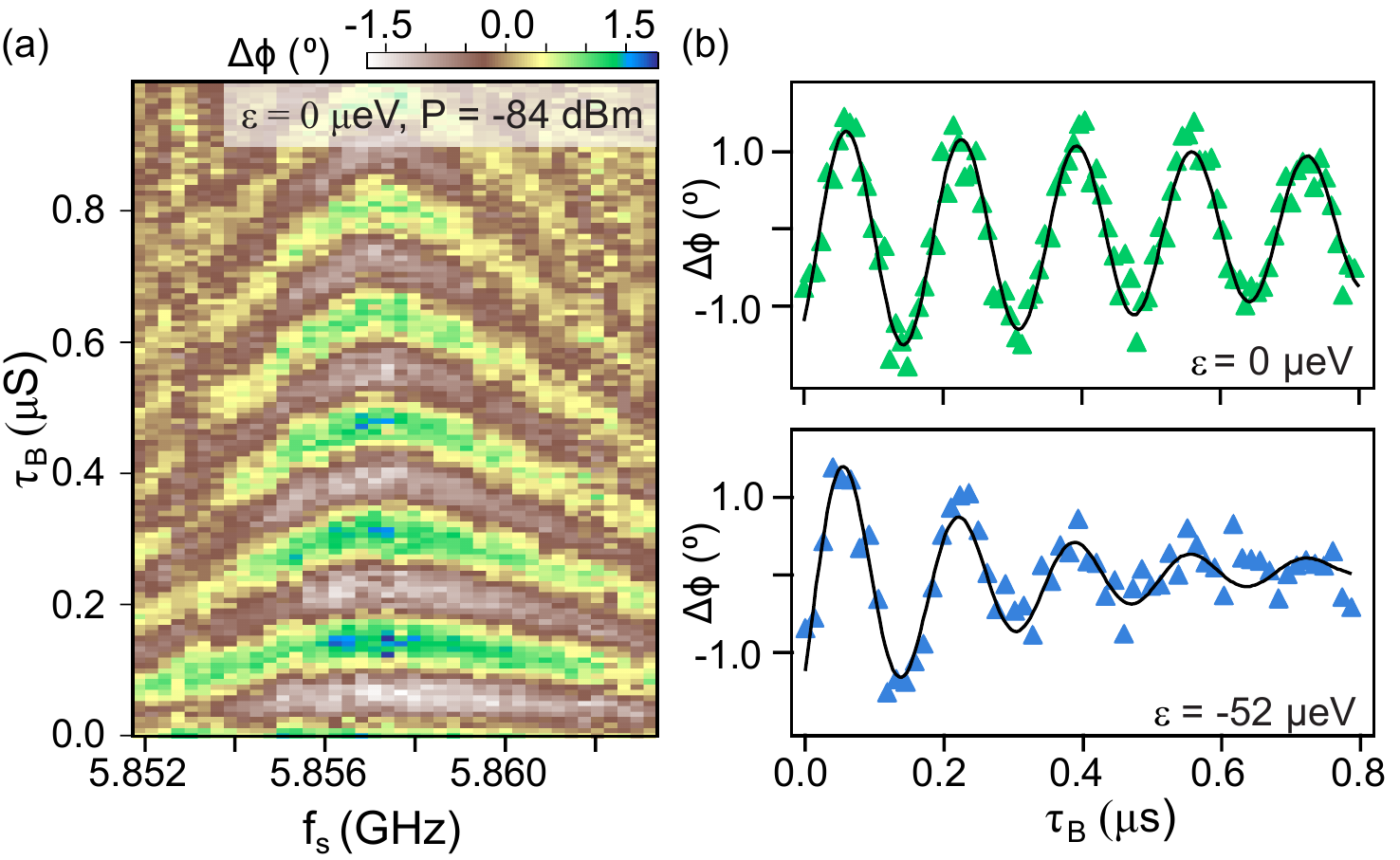}
\protect\caption{\label{fig:figure3} 
(a) Rabi chevron acquired at $\epsilon = 0$, $2t_c/h$ = 11.1~GHz, and with gate P1 driven at frequency $f_s$ for a time $\tau_B$. The cavity phase response $\Delta \phi$ is measured for 20 $\mu$s after each Rabi burst. Data are acquired with a total per point integration time of 100 ms. (b) Rabi oscillations acquired with $\epsilon = 0$ (upper panel, $P_s = -84\, \text{dBm}$) and $\epsilon = -52\, \mu\text{eV}$ (lower panel, $P_s = -77\, \text{dBm}$). The data are fit to an exponentially decaying sinusoid (solid black line). All data are acquired from DQD1.
}
\end{figure}

Having gained a quantitative understanding of how $f_{01}$ depends on $\epsilon$, we now compare EDSR in the flopping-mode and single dot regimes. By simultaneously applying a microwave burst and square pulse to gate P1, we can drive coherent spin rotations at a value of detuning set by the amplitude of the square pulse. Due to the ratio of electric dipole moments in these regimes, the power $P$ required to drive fast coherent rotations in the single dot regime is expected to be much higher. As shown in the upper panel of Fig.~\ref{fig:figure3}(b), at $\epsilon$ = 0 a Rabi frequency $f_{\rm Rabi}$ $\approx$ 6 MHz is achieved with $P$ = -84 dBm at the device. In contrast, when $\epsilon$ = -52 $\mu$eV a power of $P$ = -77 dBm is required to achieve approximately the same Rabi frequency [see Fig.~\ref{fig:figure3}(b), lower panel]. The actual power at the gate is determined by measuring the envelope of Landau-Zener-Stueckelberg interference fringes as a function of increasing $P$ \cite{Nori2010}. We fit the Rabi oscillations to an exponentially decaying sinusoid with $T_2^{\rm Rabi}$ = 1.4 $\mu$s at $\epsilon$ = 0 $\mu$eV and $T_2^{\rm Rabi}$ = 0.24 $\mu$s at $\epsilon$ = -52 $\mu$eV.

The full crossover from the single dot to the flopping-mode regime is examined over a 300 $\mu$eV range of detuning in Fig.~\ref{fig:figure4}(a). Here we plot the power required to achieve $\simeq$~6~MHz Rabi oscillations as a function of $\epsilon$. The data are nearly symmetric about $\epsilon$ = 0, as expected from the detuning symmetry of the energy levels. Most importantly, these data show that $\sim$250$\times$ less microwave power is required to achieve a $f_{Rabi}$ $\approx$ 6--8 MHz at $\epsilon$ = 0 compared to the standard single dot EDSR regime.

From theory \cite{Benito2019}, the drive power required to drive Rabi oscillations at frequency $f_{\rm Rabi}$ is given by
\begin{equation}
	 P \propto f_{\rm Rabi}^2\left(\frac{4t_c^2}{\Omega}\frac{g\mu_Bb_x}{(\Omega^2-E_z^2)}+\beta\right)^{-2}
     \, 
\end{equation}
\noindent where $\beta$ = 0.003 is a fitting parameter that accounts for single dot EDSR in the far-detuned limits ($|\epsilon| \gg 2t_c$) \cite{Golovach06,Flindt06}. We fit the data in Fig.~\ref{fig:figure4}(a) to Eq.\ (2), obtaining close agreement between theory and experiment.

\begin{figure}
\includegraphics[width=7.5cm]{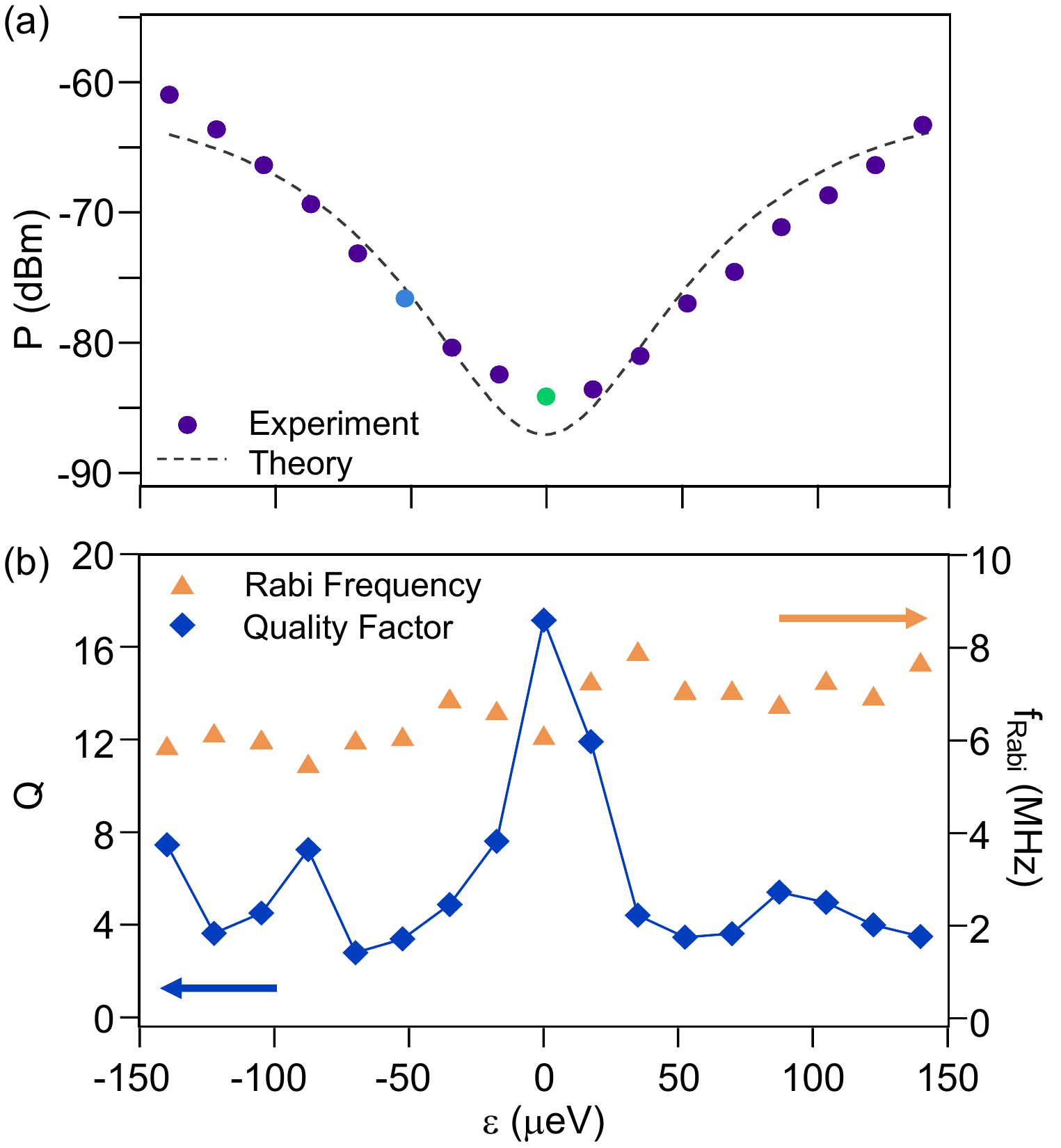}
\protect\caption{\label{fig:figure4} 
(a) On-chip power $P$ required to drive $\simeq$ 6 MHz Rabi oscillations as a function of $\epsilon$, with $2t_c/h$ = 11.1 GHz. Green (light blue) data points at $\epsilon$ = 0 ($\epsilon$ = -52 $\mu$eV) are taken from the same dataset as the top (bottom) panel in Fig.~\ref{fig:figure3}(b). (b) Quality factor $Q$ of Rabi oscillations and Rabi frequency $f_{\rm Rabi}$ plotted as a function of $\epsilon$. There is a significant improvement in $Q$ around $\epsilon$ = 0.}
\end{figure}

A charge noise sweet spot exists when $\partial E_{01}$/$\partial\epsilon = 0$ \cite{Vion02}. In the absence of a longitudinal ($b_z$) gradient the sweet spot occurs at $\epsilon$ = 0. Theory predicts that a $b_z$ gradient may shift the sweet spot to finite detuning, or if the gradient field is large enough, destroy it entirely \cite{Benito2019}. We search for evidence of a sweet spot by examining the quality factor of $\pi$-rotations $Q$ $\equiv$ $2T_2^{\rm Rabi}f_{\rm Rabi}$ as a function of detuning \cite{Yoneda2018}. At each value of detuning, a Rabi chevron is acquired with $P$ set to achieve $f_{\rm Rabi}$ $\approx$ 6 MHz, similar to Fig.~\ref{fig:figure3}(a). We take a Fourier transform of each column of the chevron and identify $f_{01}$. At this $f_{01}$, we fit the Rabi oscillations as a function of $\tau_B$ to extract $T_2^{\rm Rabi}$ and $f_{\rm Rabi}$.

The Rabi frequency and $Q$-factor are plotted as a function of $\epsilon$ in Fig.~\ref{fig:figure4}(b). At finite detuning $Q$ $\approx$ 4. We observe more than a four-fold increase in the quality factor, with $Q$ = 18 at $\epsilon$ = 0. The enhancement of $Q$ in the flopping-mode regime can be attributed to the presence of the charge noise sweet spot, which to first order decouples the spin from electrical detuning noise. While the $Q$ factors achieved here are lower than those reported elsewhere \cite{Yoneda2018, Zajac2018}, we expect that optimizing $t_c$ and fabricating devices on enriched $^{28}$Si quantum wells will further improve $Q$.

In summary, we demonstrate an efficient flopping-mode mechanism for EDSR in semiconductor DQDs. Compared to single dot EDSR, flopping-mode EDSR requires nearly three orders of magnitude less power, rendering it a valuable control technique for future spin-based quantum processors. Conveniently, the flopping-mode regime of maximal power efficiency coincides with a charge noise sweet spot, yielding a four-fold improvement in qubit quality factor. We stress that while the device studied here is embedded in a microwave cavity for readout purposes, flopping-mode EDSR could be implemented in DQDs that are read out using conventional spin-to-charge conversion \cite{Elzerman2004} or Pauli blockade \cite{Petta2005}. We anticipate that flopping-mode spin resonance will enable power-efficient single qubit control in large-scale silicon quantum processors.  

\begin{acknowledgments}
\textit{Acknowledgments---} Research sponsored by ARO grant No.\ W911NF-15-1-0149 and the Gordon and Betty Moore Foundation's EPiQS Initiative through grant GBMF4535. Devices were fabricated in the Princeton University Quantum Device Nanofabrication Laboratory.
\end{acknowledgments}

\bibliography{Croot_PRL_2019_v17_arxiv}

\end{document}